\documentclass[10pt]{article}

\usepackage{graphicx}
\usepackage{amssymb}
\usepackage{epstopdf}
\usepackage{subfigure}
\usepackage{rotating}
\usepackage{url}
\newtheorem{definition}{Definition}
\newtheorem{finding}{Finding}
\newtheorem{note}{Note}

\begin{document}

\title{Memristive excitable cellular automata}

\vspace{0.5cm}

\author{Andrew Adamatzky\\ 
University of the West of England, Bristol, UK\\
\vspace{0.5cm}
\url{andrew.adamatzky@uwe.ac.uk}\\
Leon Chua\\
University of California at Berkley, USA\\
\url{chua@EECS.Berkeley.EDU}}

\date{15 July, 2011}                                           

\maketitle

\begin{abstract}
The memristor is a device whose resistance changes depending on the polarity and magnitude of a voltage applied to the device's terminals. We design a minimalistic model of a regular network of memristors using structurally-dynamic cellular automata.  Each cell gets info about states of its closest neighbours via incoming links. A link can be one 'conductive' or 'non-conductive' states. States of every link are updated depending on states of cells the link connects. Every cell of a memristive automaton takes three states: resting, excited (analog of positive polarity) and refractory (analog of negative polarity). A cell updates its state depending on states of its closest neighbours  which are connected to the cell via 'conductive' links.  We study behaviour of memristive automata  in response to point-wise and spatially extended perturbations,  structure of localised excitations coupled with topological defects, interfacial mobile excitations and 
growth of information pathways.

\vspace{0.5cm}

\emph{Keywords:} 
memristor, cellular automaton, excitable medium

\end{abstract}

\section{Introduction}

The memristor (a passive resistor with memory) is a device whose resistance changes depending on the polarity and magnitude of a voltage applied to the device's terminals and the duration of this voltage's application. Its existence was theoretically postulated by Leon Chua in 1971 based on symmetry in integral variations of OhmÕs laws~\cite{chua:1971,chua:1976,chua:1980}. The memristor is characterised by a non-linear relationship between the charge and the flux; this relationship can be generalised to any two-terminal device in which resistance depends on the internal state of the system~\cite{chua:1976}. The memristor cannot be implemented using the three other passive circuit elements --- resistor, capacitor and inductor Ð therefore the memristor is an atomic element of electronic circuitry~\cite{chua:1971,chua:1976,chua:1980}. Using memristors one can achieve circuit functionalities that it is not possible to establish with resistors, capacitors and inductors, therefore the memristor is of great pragmatic usefulness.  The first experimental prototypes of memristors are reported in~\cite{williams:2008,erokhin:2008,yang:2008}.  Potential unique applications of memristors are in spintronic devices, ultra-dense information storage, neuromorphic circuits, and programmable electronics~\cite{strukov:2008}. 

Despite explosive growth of results in  memristor studies there is still
 a few (if any) findings on phenomenology of spatially extended non-linear media with hundreds of thousands of locally connected memristors. We attempt to fill the gap and develop a minimalistic model of a discrete 
 memristive medium. Structurally-dynamic (also called topological)  cellular automata~\cite{ilachinsky:halpern:1987}, \cite{halpern:1990}  seem to be an ideal substrate to imitate discrete memristive medium. A cellular automaton is structurally-dynamic when links between cells can be removed and reinstated depending on states of cells these links connect.  Strcturally-dynamic automata are  now proven tools to simulate physical and chemical discrete 
 spaces~\cite{rose:1994,hasslacher:1994,hillman:1998,requardt:2003,alonso-sanz:2006} and graph-rewriting media~\cite{tomita:2009}; see overview in~\cite{ilachinsky:2009}.

We must highlight that simulation of cellular automata in networks of memristors is discussed in full details 
in \cite{itoh:2009}. Itoh-Chua memristor cellular automata are automata made of memristors.  
Memristive cellular automata studied in present paper are cellular automata which exhibit, or rather roughly imitate, certain memristive properties but otherwise are classical  excitable structurally-dynamic cellular automata. 

The paper is structured as follows. Memristive automata are defined in Sect.~\ref{definition}. 
In Sect.~\ref{phenomenology} we analyse space-time dynamics of the automata in response to external perturbations. 
Anatomy of stationary oscillating localizations is presented in Sect.~\ref{localisations}. We analyse structure of 
disordered excitation worms propagating at the boundary between disorganised and ordered link domains in Sect.~\ref{interface}.  Collision-based approach to layout of conductive pathways is given in Sect.~\ref{pathways}. Some ideas of further studies are outlined in Sect.~\ref{discussion}.

\section{Memristive automaton}
\label{definition}

\begin{definition}
A memristive automaton is a structurally-dynamic excitable cellular automaton where a link connecting two cells is removed or added if one of the cells is in excited state and another cell is in refractory state.
\end{definition}

A cellular automaton $\mathcal{A}$ is an orthogonal array of uniform finite-state machines, or cells. Each cell takes finite number of states and updates its states in discrete time depending on states of its closest neighbours. All cells update their states simultaneously by the same rule. We consider eight-cell neighbourhood and three cell-states: resting $\circ$, excited $+$, and refractory $-$. Let $u(x) = \{  y: |x-y|_{L\infty}=1\}$ be a neighbourhood of cell $x$. 
A cell $x$ has a set of incoming links $\{  l_{xy}: {y \in u(x)}\}$ which take states $0$ and $1$. 
A link $l_{xy}$ is a link of excitation transfer from cell $y$ to cell $x$. 
A link in state $0$ is considered to be high-resistant, or non-conductive, and link in state $1$ low resistant, or conductive. A link-state $l^t_{xy}$ is updated depending on states of cells $x$ and $y$ at time step $t$: $l^t_{xy} = f(x^t, y^t)$. 
Resting state gives little indication of cell's previous history, therefore we will consider not resting cells 
contributing to a link state updates. When cells $x$ and $y$ are in the same state (bother cells are in state $+$ or both are in state $-$) no 'current' can flow between the cells, therefore scenarios $x^t=y^t$ are not taken into account. 
 Thus  we assume that  the only situations when $x^t, y^t \in \{+, - \}$ and $x^t \neq y^t$ may lead to changes in links conductivity:  
\begin{equation}
l_{xy}^{t+1}=
\begin{cases}
a, x^t=+ \textrm{ and } y^t=-\\
b, x^t=- \textrm{ and } y^t=+\\
l_{xy}^{t}, \textrm{ otherwise} 
\end{cases}
\label{linksequation}
\end{equation}
where $a \neq b$ and $a,b \in \{ 0, 1 \}$. Thus we consider two types of automata $\mathcal{A}^{ab}$:
$\mathcal{A}^{01}$ and $\mathcal{A}^{10}$. 

A resting cell excites ($x^t=\circ \rightarrow x^t=+$ transition) depending on number of excited neighbours. 

There are two ways to calculate a weighted sum of number of excited neighbours: 
\begin{enumerate}
\item $\Sigma_+ = \sum_{y \in u(x)}  \chi(y^t,+)$
\item $\sigma_+ = \sum_{y \in u(x)}  l_{xy}\chi(y^t,+)$,
\end{enumerate}
where $\chi(y^t,+)=1$ if $y^t=+$ and $\chi(y^t,+)=0$ otherwise.

Thus, we consider two types of memristive automata.
In automaton ${\mathcal A}_1$ a resting cell excites if $\sigma_+ > 0$. In automaton  ${\mathcal A}_2$ a 
resting cell excites if $\Sigma_+ > 1$ or $\sigma_+ > 0$.
 
 \begin{note}
Automaton ${\mathcal A}_1$ is a pre-memristive cellular automaton because it imitates only polarity (links update
(\ref{linksequation})) not voltage;  automaton ${\mathcal A}_2$ is a  memristive cellular automaton because it 
imitates both polarity (links update (\ref{linksequation})) and current intensity (use of $\Sigma_+$ and $\sigma_+$). 
\end{note}

A polarity of  a current is imitated by excitable cellular automaton using exctited $+$ and $-$ refractory states. If 
a cell $x$ is in state $-$ and a cell $y$ is in state $+$ then cell $y$ symbolises an anode, and cell $x$ a cathode. 
And we can say that a current flows from $y$ to $x$. Indeed, such an abstraction is at the edge of physical reality, however this is the only way to develop a \emph{minimal} discrete model of a memristive network.
In automaton ${\mathcal A}_2$  the condition  $\Sigma_+ > 1$  symbolises propagation of a 
 high intensity current along all links, including links non-conductive for a low intensity current. 
 This high intensity current in ${\mathcal A}_2$ resets conductivity of the links and also states of cells. 
 The condition  $\sigma_+ > 0$ reflects propagation of a low intensity current along conductive links. 
 The  current  of low intensity does not affect states of links but only states of cells.

\subsection{Experiments}

We experiment with $300 \times 300$ cell automaton arrays, with  non-periodic absorbing boundaries.
We conduct experiments for two initial conditions on links' 'conductivity' ---
\begin{itemize}
\item $L_1$-condition: all links are conductive (for every cell $x$ and its neighbour $y$ $l^0_{xy}=1$), and 
\item $L_0$-condition: all links are non-conductive  (for every cell $x$ and its neighbour $y$ $l^0_{xy}=0$). 
\end{itemize}

While testing automata's response to external excitation we use point-wise and spatially extended stimulations. 
By point-wise stimulation we mean excitation of a single cell ($\mathcal{A}_1$) or a couple of cells  ($\mathcal{A}_2$)
of resting automata. These are minimal excitations to start propagating activity.

Let $D$-disc be a set of cells which lie at distance not more than $r$ from the array centre $(n/2, n/2)$, $n=300, 
r=n/3)$. When undertaking $D$-stimulation we assign excited state $+$ to a cell of $D$ with probability 0.05. In some cases we apply $D$-stimulation twice as follows. An automaton starts in $L_1$- or $L_2$-condition, we apply $D$-stimulation first time (we call it $E_1$-excitation) and wait till excitation waves propagate beyond boundaries of the array or a quasi-stationary structure is formed. After this transient period we apply $D$-stimulation again ($E_2$-excitation) without resetting links states.  

Space-time dynamics of automata is illustrated by configurations of excitations and dynamics of link conductivity is shown  either explicitly by arrows (in small configuration) or via grey-scale representation of  cells' in-degrees: cell $x$ is represented by pixel with grey-value $32 \cdot \sum_{y \in u(x)} $.   Despite not representing exact configuration of local links,  in-degrees give us a rough indicator of spatial distribution of conductivity in the medium. 
The higher is the in-degree at a given point, the higher is the conductivity at this point.

\section{Phenomenology}
\label{phenomenology}

\begin{finding}
A point-wise stimulation of automaton $\mathcal{A}_2$ leads to a persistent excitation, while automaton
$\mathcal{A}_1$ returns to a resting state.   
\end{finding}

\begin{figure}[!htb]
\centering
\subfigure[$t=1$]{\includegraphics[width=0.32\textwidth]{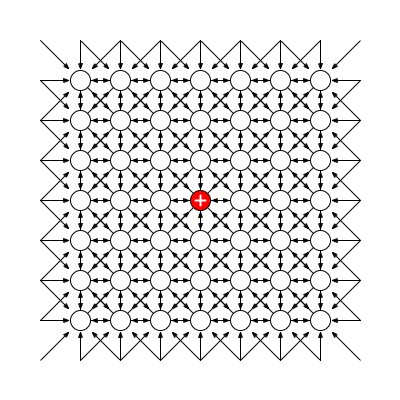}}
\subfigure[$t=2$]{\includegraphics[width=0.32\textwidth]{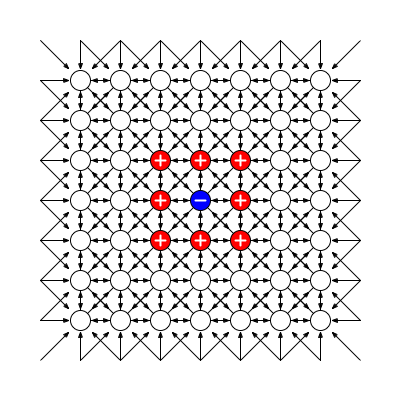}}
\subfigure[$t=3$]{\includegraphics[width=0.32\textwidth]{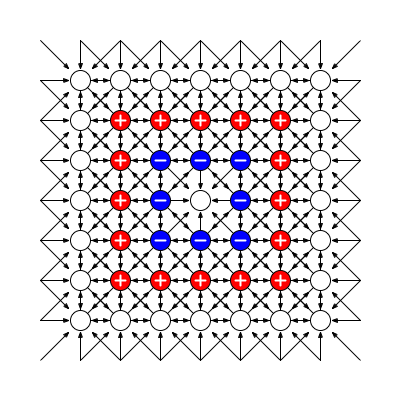}}
\subfigure[$t=4$]{\includegraphics[width=0.32\textwidth]{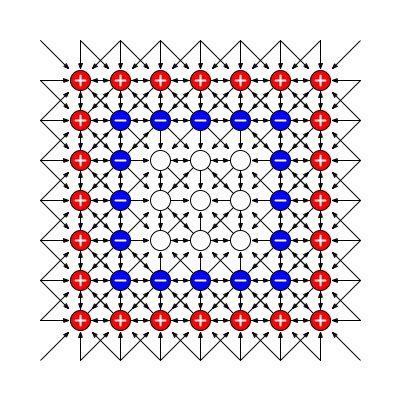}}
\subfigure[$t=5$]{\includegraphics[width=0.32\textwidth]{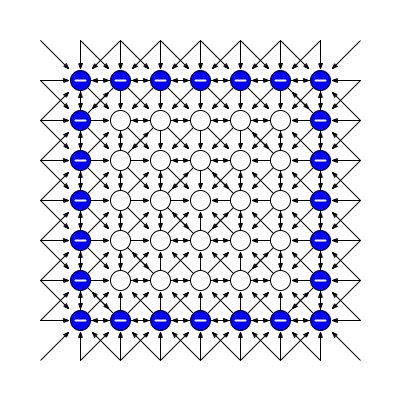}}
\subfigure[$t=6$]{\includegraphics[width=0.32\textwidth]{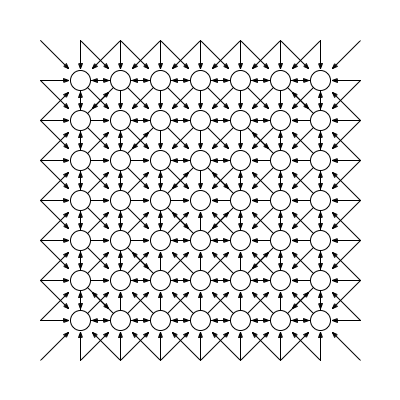}}
\caption{Snapshots of excitation and links dynamics in automaton  $\mathcal{A}_1^{01}$. In initial configuration
($t=1$) one cell is excited others are resting. Arrow from cell $x$ to cell $y$ means $l_{xy}=1$. 
Excited cells are shown by red discs with sign '+', refractory cells by blue discs with sign '-', resting cells are blank.
Links at the edges of cellular array corresponds to an absorbing boundary cells, which are always in resting state independently on states of their neighbours.
}
\label{singleexcitationtopology}
\end{figure}

\begin{figure}[!htb]
\centering
\subfigure[$t=1$]{\includegraphics[width=0.32\textwidth]{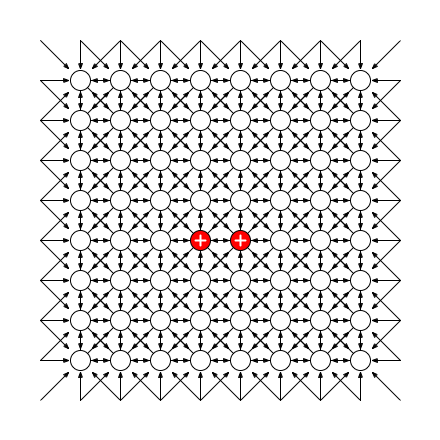}}
\subfigure[$t=2$]{\includegraphics[width=0.32\textwidth]{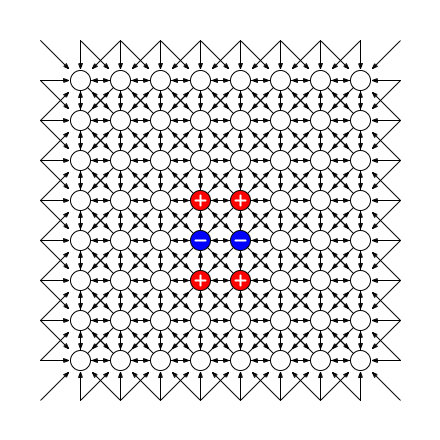}}\\
\subfigure[$t=3$]{\includegraphics[width=0.32\textwidth]{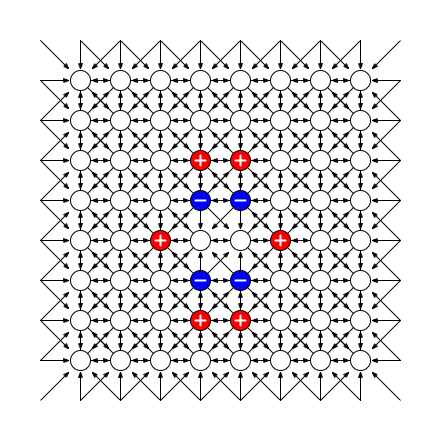}}
\subfigure[$t=4$]{\includegraphics[width=0.32\textwidth]{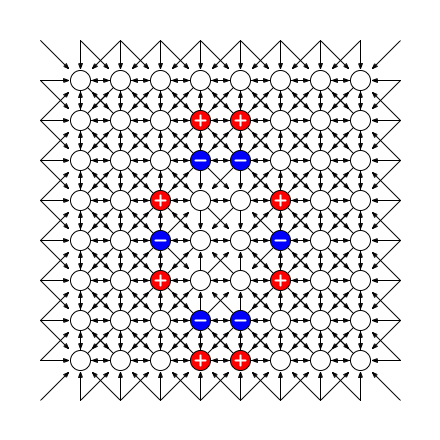}}
\caption{Snapshots of excitation and links dynamics in automaton  $\mathcal{A}_2^{01}$. In initial configuration
 two cells (minimum size of non-dying excitation for $\mathcal{A}_2$) are excited others are resting. Arrow from cell $x$ to cell $y$ means $l_{xy}=1$. Excited cells are shown by red discs with sign '+', refractory cells by blue discs with sign '-', resting cells are blank}
\label{towcellexcitationtopologyA2(01)}
\end{figure}

A single-cell excitation of resting automaton $\mathcal{A}_1$  (Fig.~\ref{singleexcitationtopology}) or two-cell excitation of resting automaton  $\mathcal{A}_2$ (Fig.~\ref{towcellexcitationtopologyA2(01)}) in $L_1$ initial conditions lead to formation of  a 'classical' excitation wave-front. The wave-front propagates omni-directionally
away from the initial perturbation site and updates states of links it is passing through.   

Links leading from cells
to the neighbours they excited are made non-conductive in development of $\mathcal{A}_i^{01}$ 
(Figs.~\ref{singleexcitationtopology} and ~\ref{towcellexcitationtopologyA2(01)}); or we can say that links corresponding to normal vectors of propagating wave-front are made non-conductive. Cell excited at time step $t=1$ becomes isolated. The situation is similar in development of automata $\mathcal{A}_1^{10}$ and $\mathcal{A}_2^{10}$ with the only difference that links connecting cells which are excited at time step $t$ to cells they have been excited by are removed. 

\begin{figure}[!htb]
\centering
\subfigure[$t=1$]{\includegraphics[width=0.32\textwidth]{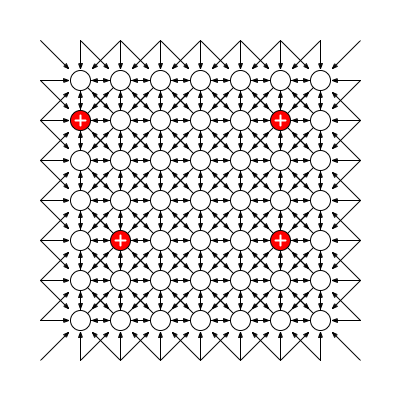}}
\subfigure[$t=2$]{\includegraphics[width=0.32\textwidth]{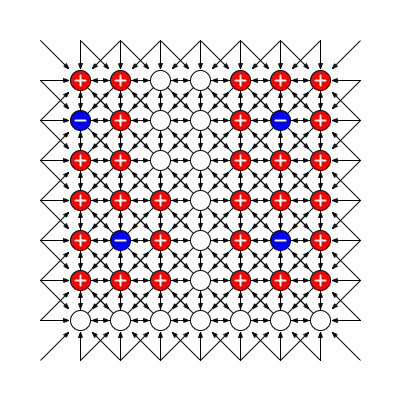}}
\subfigure[$t=3$]{\includegraphics[width=0.32\textwidth]{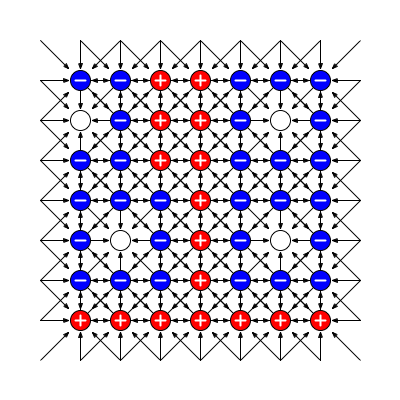}}
\subfigure[$t=4$]{\includegraphics[width=0.32\textwidth]{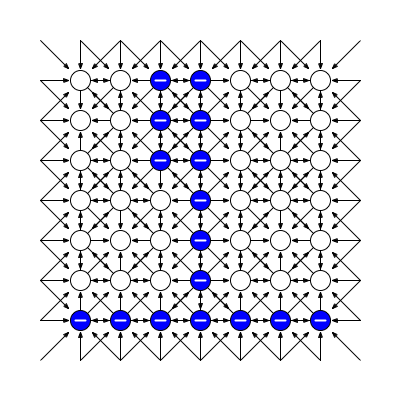}}
\subfigure[$t=5$]{\includegraphics[width=0.32\textwidth]{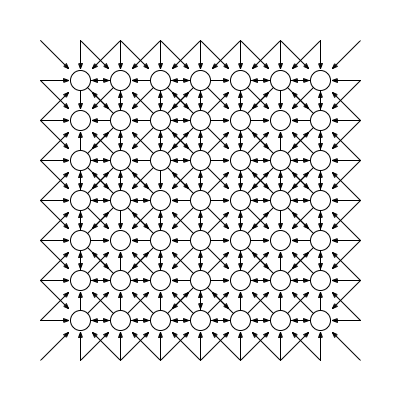}}
\caption{Snapshots of excitation and links dynamics in automaton  $\mathcal{A}_1^{01}$. In initial configuration
 several cells (those in '+'-state at $t=1$) are excited others are resting. Arrow from cell $x$ to cell $y$ means $l_{xy}=1$. }
\label{boundaryformationinA1(01)}
\end{figure}

In summary, in automata  $\mathcal{A}_i^{01}$ links associated with forward propagation of perturbation are made non-conductive, and in $\mathcal{A}_i^{10}$  links associated with backward propagation are  made non-conductive. Excitation wave-front travelling from a single stimulation site forms a domain of co-aligned links. Excitation waves initiated in different cells collide and merge.  Boundaries between domains formed by different fronts are represented by distinctive configurations of links (Fig.~\ref{boundaryformationinA1(01)}).

Automaton $\mathcal{A}_1$ is non-excitable in $L_0$-conditions because no excitation can propagate along non-conductive links.  An outcome of two-site excitation of resting automaton $\mathcal{A}_2$ in $L_0$-condition   depends on configuration of the initial excitation. 

\begin{figure}[!htb]
\centering
\subfigure[$t=1$]{\includegraphics[width=0.15\textwidth]{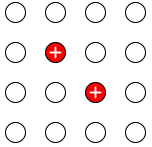}}
\subfigure[$t=2$]{\includegraphics[width=0.15\textwidth]{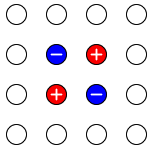}}
\subfigure[$t=3$]{\includegraphics[width=0.15\textwidth]{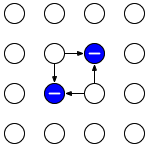}}
\subfigure[$t=4$]{\includegraphics[width=0.15\textwidth]{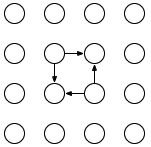}}
\caption{Snapshots of excitation and links dynamics in automaton  $\mathcal{A}_2^{01}$,  $L_0$-start. 
In initial configuration two cells (minimum size of non-dying excitation in $\mathcal{A}$) are excited others are resting. Arrow from cell $x$ to cell $y$ means $l_{xy}=1$. }
\label{TwoDiagonalCellsExcitationTopologyA2(01)NoLinksStart}
\end{figure}

Let  two diagonal neighbours (north-west and south-east) be excited (Fig.~\ref{TwoDiagonalCellsExcitationTopologyA2(01)NoLinksStart}a) at time step $t$. These two cells transfer excitation to their two neighbours  (north-east and south-west) as shown in Fig.~\ref{TwoDiagonalCellsExcitationTopologyA2(01)NoLinksStart}b. Only links from north-west cell to north-east and south-west, and south-east to south-west and north-east are formed  (Fig.~\ref{TwoDiagonalCellsExcitationTopologyA2(01)NoLinksStart}c) and excitation becomes extinguished  (Fig.~\ref{TwoDiagonalCellsExcitationTopologyA2(01)NoLinksStart}d). 

\begin{figure}[!htb]
\centering
\subfigure[$t=1$]{\includegraphics[width=0.32\textwidth]{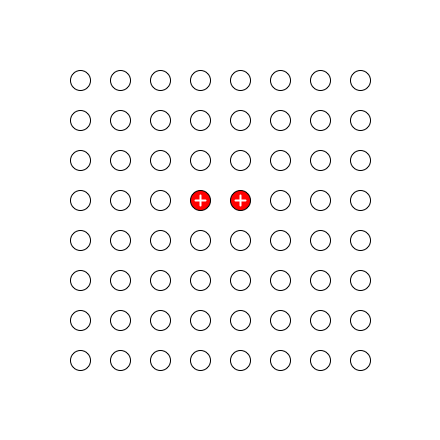}}
\subfigure[$t=2$]{\includegraphics[width=0.32\textwidth]{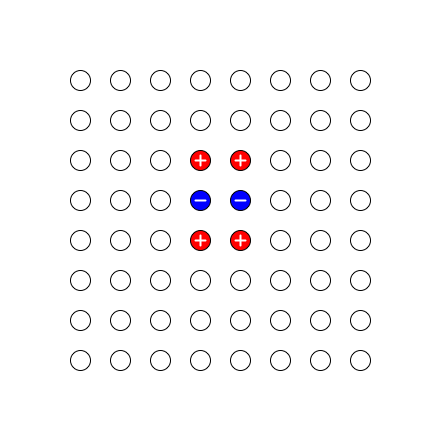}}
\subfigure[$t=3$]{\includegraphics[width=0.32\textwidth]{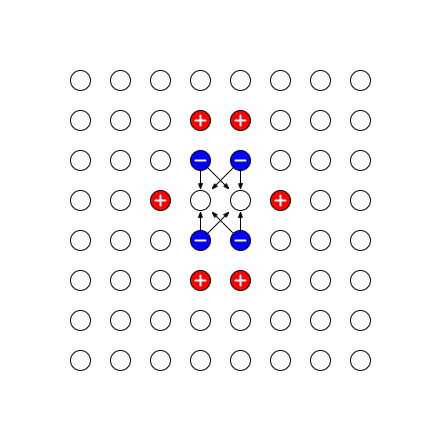}}
\subfigure[$t=4$]{\includegraphics[width=0.32\textwidth]{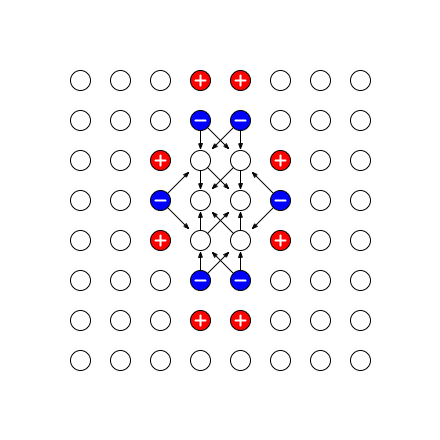}}
\subfigure[$t=5$]{\includegraphics[width=0.32\textwidth]{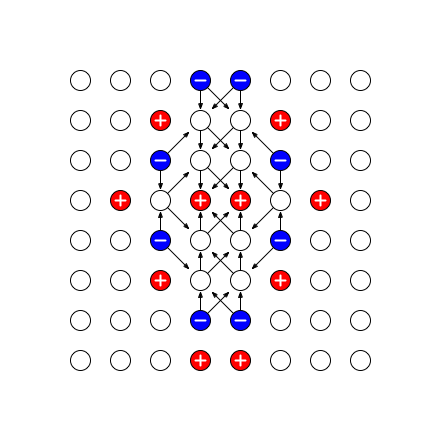}}
\caption{Snapshots of excitation and links dynamics in automaton  $\mathcal{A}_2^{10}$, $L_0$-start. 
In initial configuration two cells (minimum size of non-dying excitation) are excited others are resting. 
Arrow from cell $x$ to cell $y$ means $l_{xy}=1$. }
\label{TwHorizontalCellsExictationTopologyA2(10)NoLinksStart}
\end{figure}

Let excited cells be neighbours at the same row of cells (Fig.~\ref{TwHorizontalCellsExictationTopologyA2(10)NoLinksStart}a). If they are excited at time step $t$ then their north and south neighbours are excited due to 
condition $\Sigma_+ > 1$ taking place  (Fig.~\ref{TwHorizontalCellsExictationTopologyA2(10)NoLinksStart}b).
The localised (two-cell size) excitations propagate north and south  (Fig.~\ref{TwHorizontalCellsExictationTopologyA2(10)NoLinksStart}c) and make links pointing backwards (towards source of initial excitation) conductive. At the second iteration cell lying east and west of initially perturbed cells becomes excited (Fig.~\ref{TwHorizontalCellsExictationTopologyA2(10)NoLinksStart}c). By that initially perturbed cells return to resting state
(Fig.~\ref{TwHorizontalCellsExictationTopologyA2(10)NoLinksStart}d) and thus they become excited again.  A growing pattern of recurrent excitation fills the lattice (Fig.~\ref{TwHorizontalCellsExictationTopologyA2(10)NoLinksStart}de).

\begin{finding}
Repeated stimulation of memristive automata in a spatially-extended domain leads to formation of either 
disorganised activity domain emitting target waves of excitation 
or a sparse configuration of stationary oscillating localizations.  
\end{finding}

\begin{figure}[!htb]
\centering
\includegraphics[width=0.8\textwidth]{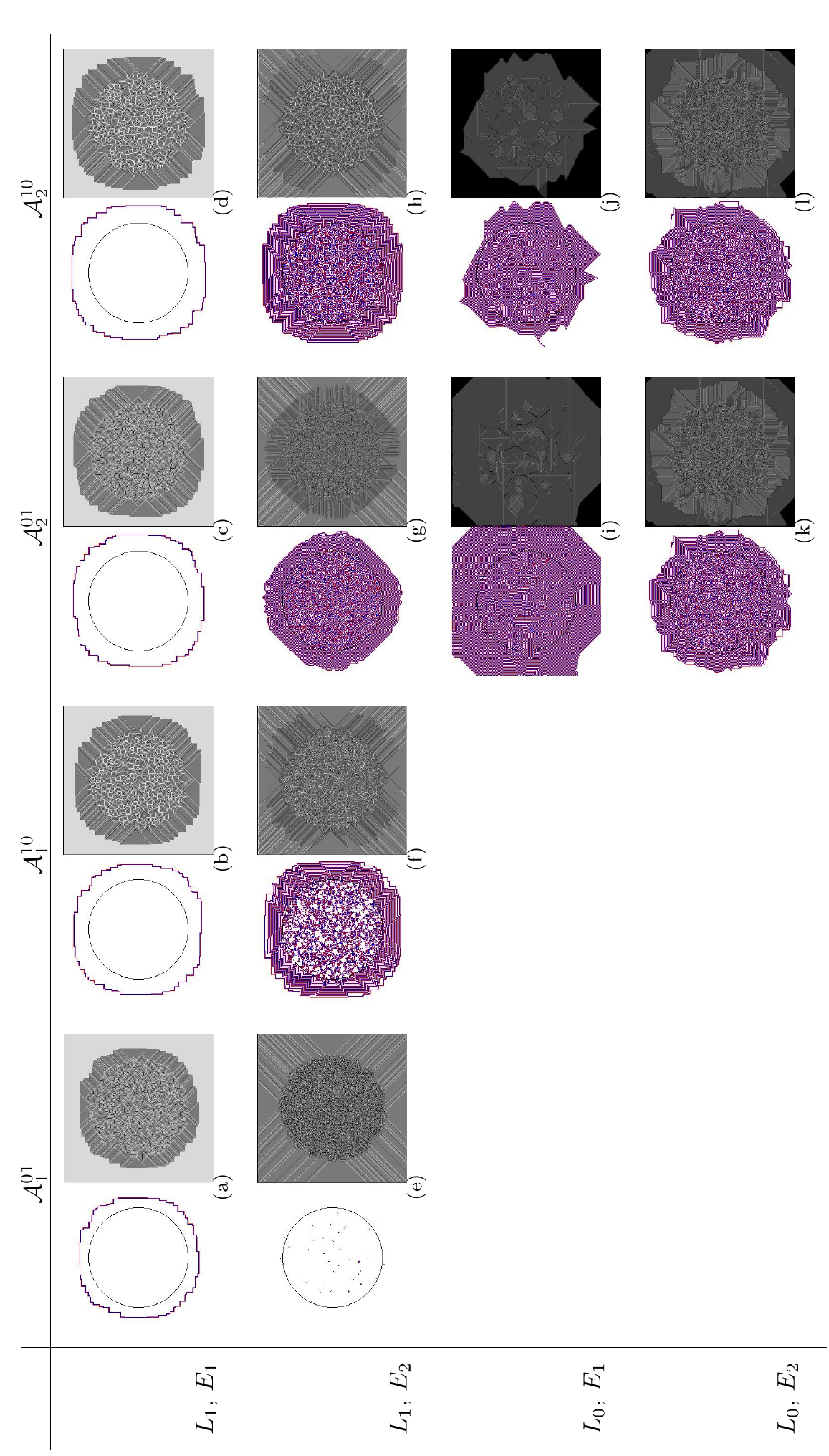}
\caption{Example of automaton behaviour in response to $D$-stimulation.  Boundaries of disc $D$  are shown as circles. Topology of links is represented by grey-levels of  cells' in-degrees.}
\label{ExamplesOfRandomExcitation}
\end{figure}

Examples of  excitation dynamics in response to $E_1$- and $E_2$-excitations  are shown in Fig.~\ref{ExamplesOfRandomExcitation}. Each singular perturbation in the first stimulation  
of automata being in $L_1$-condition leads to propagation of excitation. The waves merge 
into a single wave and disappear beyond lattice boundary.
Merging of waves outside $D$ into a single wave is reflected in cells' in-degree distribution in 
Fig.~\ref{outsidediscdegrees}.

Automata' behaviour become different after second $D$-stimulation ($E_2$-excitation). They all exhibit quasi-chaotic dynamic inside boundaries of disc $D$  (shown by circle in Fig.~\ref{ExamplesOfRandomExcitation}); cell in-degrees' distributions of the automata are similar (Fig.~\ref{insidediscdegrees}). 
However the excitation dynamics in automaton $\mathcal{A}_1^{01}$ is reduced (after long transient period) to stationary oscillating localizations while in automata   $\mathcal{A}_1^{10}$, $\mathcal{A}_2^{01}$, and $\mathcal{A}_2^{10}$
excitations outside boundaries of initial stimulation merge into target waves 
(Fig.~\ref{ExamplesOfRandomExcitation} and Fig.~\ref{outsidediscdegrees}). Oscillating localizations developed in 
$\mathcal{A}_1^{01}$ stay inside the boundaries of $D$.

Random excitation is extinguished immediately in automata $\mathcal{A}_1$ being in $L_0$-condition of total non-conductivity. When automaton $\mathcal{A}_2$ in $L_0$-condition is excited, the quasi-random excitation activity persists inside  boundaries of $D$ while omni-directional waves are formed outside $D$. Patterns of activity are not changed significantly after second random perturbation, $E_1$-excitation (Fig.~\ref{ExamplesOfRandomExcitation}).

 If there are both excited and refractory states in the external stimulation domain the developments are almost the same but $\mathcal{A}_1^{10}$ and $\mathcal{A}_2^{10}$  shows persistent excitation activity already at the first stimulation.

\clearpage

\section{Oscillating localisations}
\label{localisations}

\begin{finding}
$E_2$-excitation of $\mathcal{A}_1^{01}$ leads to formation  of excitation wave-fragments trapped in a structurally defined domains. 
\end{finding}

\begin{figure}[!htb]
\centering
\subfigure[]{\includegraphics[width=0.49\textwidth]{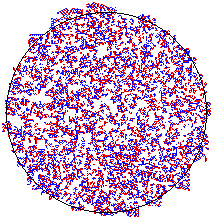}}
\subfigure[]{\includegraphics[width=0.49\textwidth]{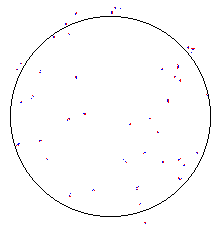}}
\subfigure[]{\includegraphics[width=0.65\textwidth]{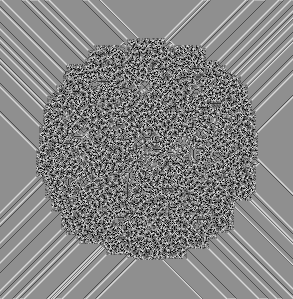}}
\caption{Formation of localised excitations in $\mathcal{A}_1^{01}$,  $L_1$-start. A localised 
excitation occupies a fixed size domain, the excitation may change its size periodically but 
never expands more then the certain fixed size.
(a)~configuration of automaton after $E_2$-excitation, boundary of $D$ is shown by black
circle; (b)~the same automaton after transient period, only few oscillating localizations sustain, (c)~links' 
conductivity presented via grey-values of cells' in-degrees. }
\label{largeconfiguration}
\end{figure}

\begin{figure}[!htb]
\centering
$\frac{\includegraphics[width=0.24\textwidth]{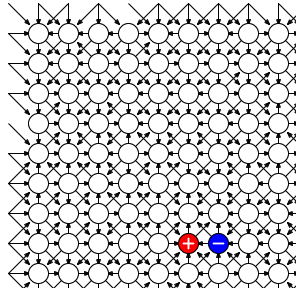}}{t=0}$
$\frac{\includegraphics[width=0.24\textwidth]{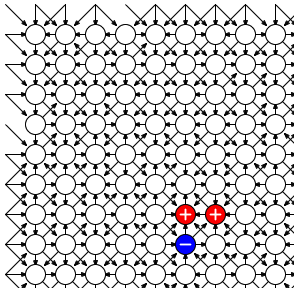}}{t=1}$
$\frac{\includegraphics[width=0.24\textwidth]{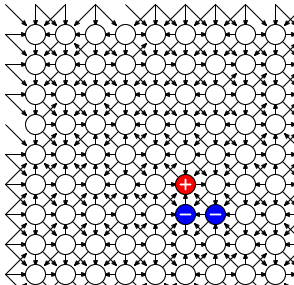}}{t=2}$
$\frac{\includegraphics[width=0.24\textwidth]{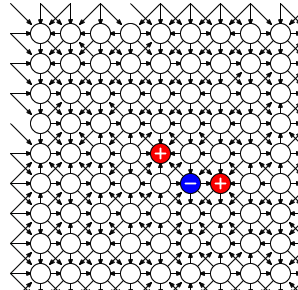}}{t=3}$
$\frac{\includegraphics[width=0.24\textwidth]{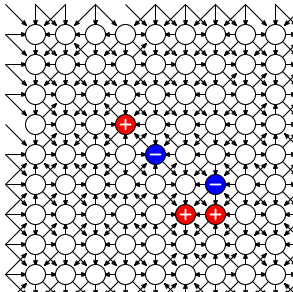}}{t=4}$
$\frac{\includegraphics[width=0.24\textwidth]{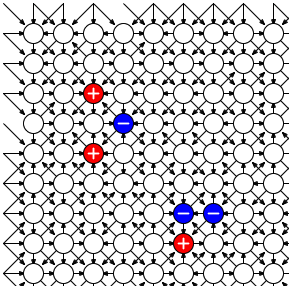}}{t=5}$
$\frac{\includegraphics[width=0.24\textwidth]{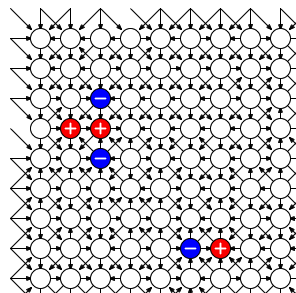}}{t=6}$
$\frac{\includegraphics[width=0.24\textwidth]{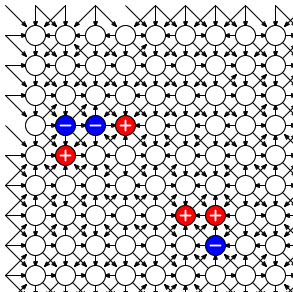}}{t=7}$
$\frac{\includegraphics[width=0.24\textwidth]{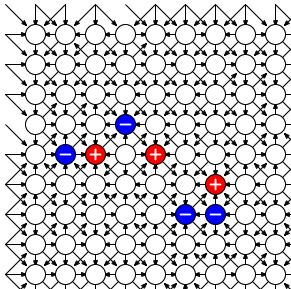}}{t=8}$
$\frac{\includegraphics[width=0.24\textwidth]{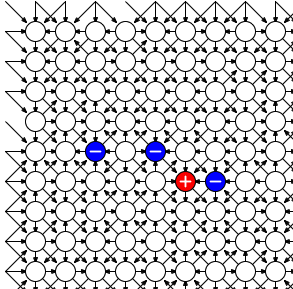}}{t=9}$
$\frac{\includegraphics[width=0.24\textwidth]{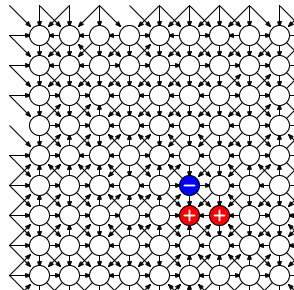}}{t=10}$\\
$\frac{\includegraphics[width=0.24\textwidth]{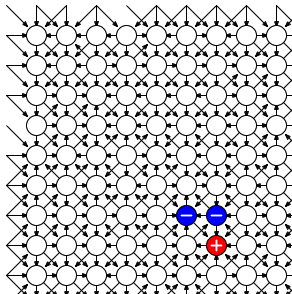}}{t=11}$
$\frac{\includegraphics[width=0.24\textwidth]{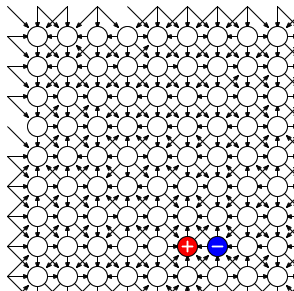}}{t=12}$
\caption{Example of oscillator $O_1$ in $\mathcal{A}_1^{01}$. }
\label{oscillator1}
\end{figure}

\begin{figure}[!htb]
\centering
$\frac{\includegraphics[width=0.20\textwidth]{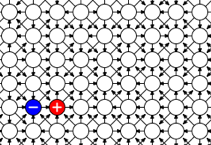}}{t=0}$
$\frac{\includegraphics[width=0.20\textwidth]{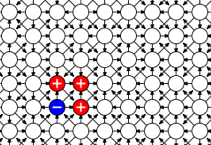}}{t=1}$
$\frac{\includegraphics[width=0.20\textwidth]{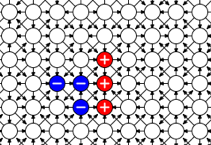}}{t=2}$
$\frac{\includegraphics[width=0.20\textwidth]{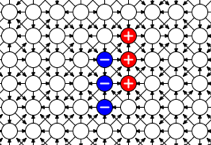}}{t=3}$
$\frac{\includegraphics[width=0.20\textwidth]{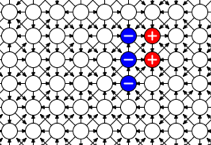}}{t=4}$
$\frac{\includegraphics[width=0.20\textwidth]{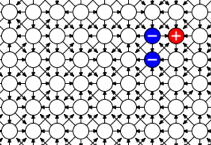}}{t=5}$
$\frac{\includegraphics[width=0.20\textwidth]{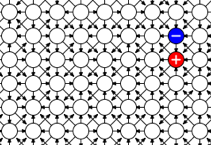}}{t=6}$
$\frac{\includegraphics[width=0.20\textwidth]{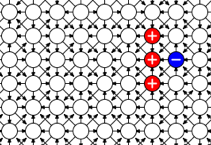}}{t=7}$
$\frac{\includegraphics[width=0.20\textwidth]{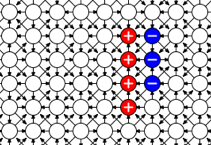}}{t=8}$
$\frac{\includegraphics[width=0.20\textwidth]{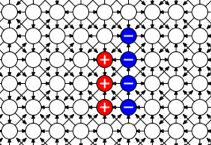}}{t=9}$
$\frac{\includegraphics[width=0.20\textwidth]{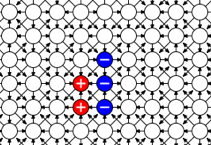}}{t=10}$
$\frac{\includegraphics[width=0.20\textwidth]{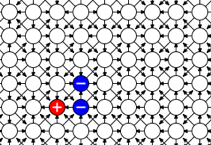}}{t=11}$
$\frac{\includegraphics[width=0.20\textwidth]{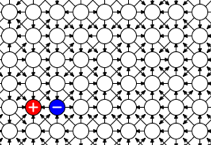}}{t=12}$
$\frac{\includegraphics[width=0.20\textwidth]{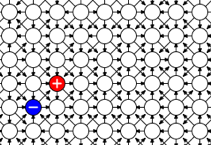}}{t=13}$
$\frac{\includegraphics[width=0.20\textwidth]{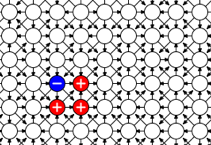}}{t=14}$
$\frac{\includegraphics[width=0.20\textwidth]{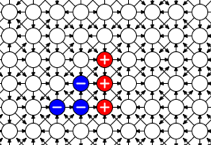}}{t=15}$
$\frac{\includegraphics[width=0.20\textwidth]{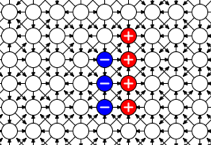}}{t=16}$
$\frac{\includegraphics[width=0.20\textwidth]{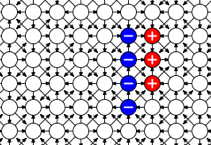}}{t=17}$
$\frac{\includegraphics[width=0.20\textwidth]{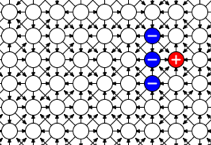}}{t=18}$
$\frac{\includegraphics[width=0.20\textwidth]{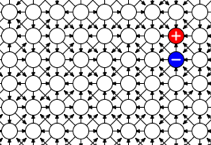}}{t=19}$
$\frac{\includegraphics[width=0.20\textwidth]{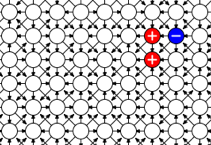}}{t=20}$
$\frac{\includegraphics[width=0.20\textwidth]{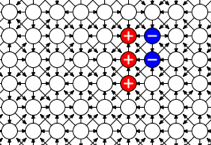}}{t=21}$
$\frac{\includegraphics[width=0.20\textwidth]{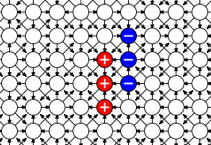}}{t=22}$
$\frac{\includegraphics[width=0.20\textwidth]{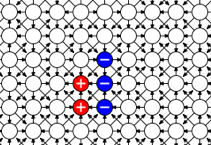}}{t=23}$
$\frac{\includegraphics[width=0.20\textwidth]{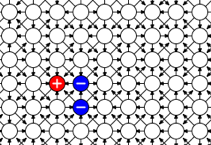}}{t=24}$
$\frac{\includegraphics[width=0.20\textwidth]{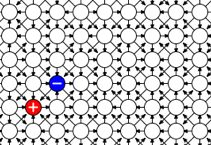}}{t=25}$
$\frac{\includegraphics[width=0.20\textwidth]{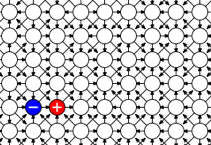}}{t=26}$
\caption{Example of oscillator $O_2$ in  $\mathcal{A}_1^{01}$. }
\label{oscillator3}
\end{figure}

\begin{figure}[!htb]
\begin{center}
\begin{tabular}{l|cc}
Characteristic 		& $O_1$ & $O_2$ \\\hline
Period           		&   12         &  26       \\
Minimum mass           	&     2       &   2      \\
Maximum mass           	&      7      &   7      \\
Minimum size           	&      2      &  2       \\
Maximum size           	&      36      &   9       \\
Minimum density           	&       1     &   1      \\
Maximum density           	&      $ \frac{1}{6}$     & $ \frac{1}{2}$        \\
\end{tabular}
\end{center}
\caption{Characteristics of commonly found oscillators $O_1$ and $O_2$ in $\mathcal{A}_1^{01}$.}
\label{oscillatorscharacteristics}
\end{figure}%

$E_2$-excitation of  $\mathcal{A}_1^{01}$ leads to formation of sparsely distributed 
localised oscillating excitations, or oscillators (Fig.~\ref{largeconfiguration}). An oscillating localisation (oscillator) usually 
consist of one or two  mobile localizations which shuffle inside a small compact domain of the automaton array. 
This micro-wave is updating states of links and thus influencing its own behaviour.  
$\mathcal{A}_1^{01}$, after $E_2$-stimulation and formation of oscillating localizations, is characterised by a smooth balanced distribution of cells in-degrees (Fig.~\ref{insidediscdegrees}, table entry $L_1$, $E_2$).

Examples of two most commonly found oscillators --- $O_1$ and $O_2$ --- are shown in 
Figs.~\ref{oscillator1} and \ref{oscillator3}, and their   characteristics 
in Fig.~\ref{oscillatorscharacteristics}. Both oscillators have exactly the same minimum and maximum masses
 (measured as a sum of cells in excited and refractory states). Oscillator $O_2$ has much longer period than $O_1$ and larger maximum density. Oscillator $O_1$  spans large space during its transformation cycle, it occupies 
a sub-array of $6 \times 6$ cells when in its largest form.  

In its minimal form $O_1$ consists of two cells: one cell is in state $+$ another in state $-$ 
(Fig.~\ref{oscillator1}, $t=0$). At next two steps of $O_1$'s transformations a small localised excitation is formed. It propagates north
(Fig.~\ref{oscillator1}, $t=1,2,3$).  At fourth step of oscillator $O_1$'s transformation the travelling localised excitation splits  into two excitations: one travels north-west, another south (Fig.~\ref{oscillator1}, $t=4,5$). The localisation travelling south returns to exact position of the first step excitation, just with swapped excited and refractory states 
  (Fig.~\ref{oscillator1}, compare $t=6$ with $t=0$) and then repeats a cycle of transformations $t=1,2$ (compare 
  $t=1$ with $t=7$ an $t=2$ with $t=8$). The excitation travelling north-west become extinguished ($t=8,9$). 

A couple $(-+)$ is a minimal configuration of oscillator $O_2$ (Fig.~\ref{oscillator3}, $t=0$). 
When the localisation starts it is development in configuration $(-+)$, and configuration of links'
conductivity as shown in Fig.~\ref{oscillator3}, it is transformed into excitation wave-fragment propagating east and north-east  (Fig.~\ref{oscillator3}, $t=1,2,3$). At fourth step of oscillator's transformation the wave-fragment shrinks
and by step $t=6$ the oscillator repeats its original state $(-+)$ yet rotated by $90^o$ clockwise. New 
excitation wave-fragment emerges and propagates west and south-west (Fig.~\ref{oscillator3}, $t=7 \ldots 10$).
The fragment contracts to configuration $(+-)$ by step $t=12$. Then development and transformation of 
excitation wave-fragments is repeated (Fig.~\ref{oscillator3}, $t=13, \cdots, 23$). The localisation returns 
to its original configuration $(-+)$ at $t=26$. 

Most oscillating localizations observed in experiments with $\mathcal{A}_1^{01}$ have very long periods. This is because the oscillators' behaviour is determined not only by cell-state transitions rules, as in classical cellular automata, but also by topology of links modified by repeated random stimulation and dynamics of the links affected by 
oscillators themselves.

\clearpage

\section{Dynamics of excitation on interfaces}
\label{interface}

\begin{finding}
Automaton $\mathcal{A}_1^{01}$ exhibits localizations propagating along the boundary of disordered and ordered
conductivity domains.
\end{finding}

\begin{figure}[!htb]
\centering
\subfigure[$t$]{\includegraphics[width=0.4\textwidth]{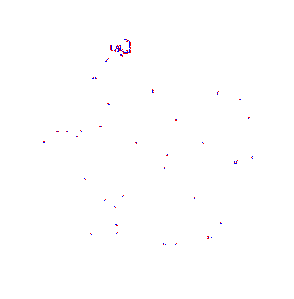}}
\subfigure[$t+200$]{\includegraphics[width=0.4\textwidth]{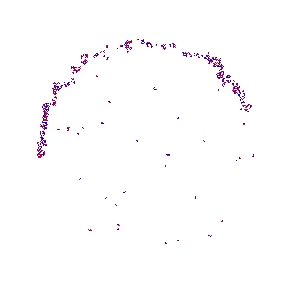}}
\subfigure[$t+300$]{\includegraphics[width=0.4\textwidth]{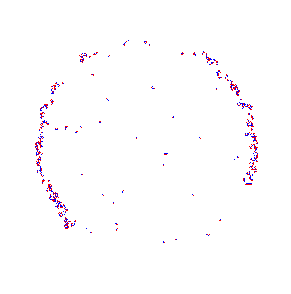}}
\subfigure[$t+500$]{\includegraphics[width=0.4\textwidth]{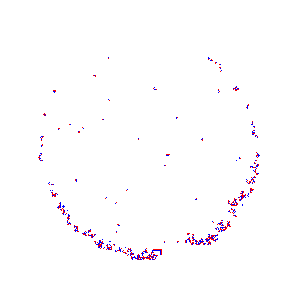}}
\subfigure[$t+600$]{\includegraphics[width=0.4\textwidth]{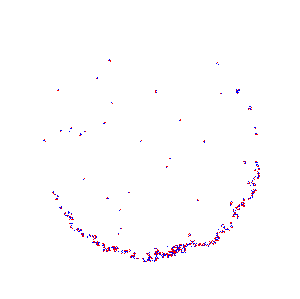}}
\subfigure[$t+800$]{\includegraphics[width=0.4\textwidth]{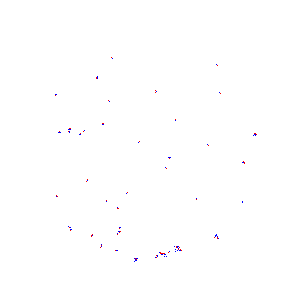}}
\caption{ Propagation of localised excitations at the interface between disordered 
and ordered conductivity domains,  $\mathcal{A}_1^{01}$ automaton, $L_1$-start. }
\label{propagationatinterface}
\end{figure}

As previously discussed, after $E_2$-excitation of  $\mathcal{A}_1^{01}$ the automaton 
becomes separated into two domains. One domain has a disorganised appearance, it lies inside boundaries 
of $D$. This domain is characterised by quasi-random discoidal configuration of cell 
in-degrees (Fig.~\ref{largeconfiguration}c). Second domain, cells lying outside $D$, 
represents ordered configurations of links (Fig.~\ref{largeconfiguration}c). The links are co-alined 
during target wave propagation outward $D$.  See Figs.~\ref{insidediscdegrees} 
and~\ref{outsidediscdegrees} for distribution of cell in-degrees in disordered and ordered domaines.

In automaton $\mathcal{A}_1^{01}$ a link from cell $y$ to cell $x$ becomes non-conductive at time step $t+1$ if $x^t=+$ and $y^t=-$. Thus target waves travel ouside of $D$ form non-conductive pathways. If we stimulate automaton in these domains of ordered links no distal propagation of excitation will be observed,  however, a worm 
of disordered excitations is formed at the boundary between disordered and ordered configurations of links.

\begin{figure}[!htb]
\centering
\includegraphics[width=0.6\textwidth]{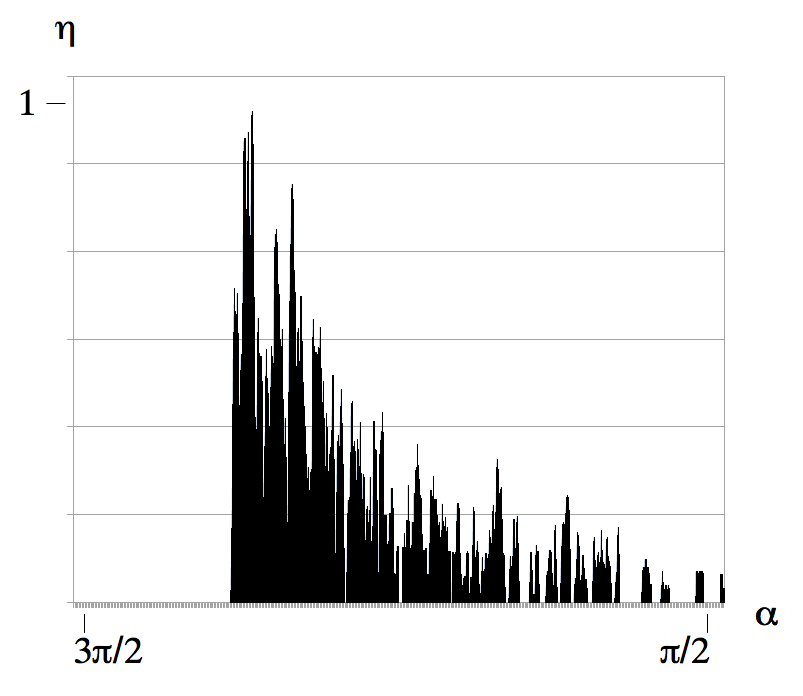}
\caption{Illustrative profile of a worm excitation 
propagating anti-clockwise in Fig.~\ref{propagationatinterface}. $\eta(\alpha)$ is a normalised number of cells 
in non-resting states lying at  distance $[n/3-10, n/3+10]$ and angle $\alpha$ from the centre of disc $D$.}
\label{wavefront}
\end{figure}

When configuration of oscillating localizations becomes stationary (in a sense that the localizations 
do not travel outside their fixed domains) we apply a few-cells wide excitation at the boundary of  $D$ (Fig.~\ref{propagationatinterface}a), north-north-west localised excitation, which exceeds in size existing oscillating localizations). An extended cluster, or a worm (a group of excited and refractory states with a preferable direction of growth), 
of excited and refractory cells is formed (Fig.~\ref{propagationatinterface}b). It expands along the boundary between domains. Two heads of the emerging worm --- one head moves clockwise and another anticlockwise --- are clearly visible in  (Fig.~\ref{propagationatinterface}). At some point the worm undergoes sub-division  
(Fig.~\ref{propagationatinterface}c) and it breaks into two independent worms (Fig.~\ref{propagationatinterface}d). Each new worm has an extended head and gradually thinning tail. Thickness of a worm, 
measured in a number of non-resting cells along the worm body, is shown in Fig.~\ref{wavefront}.

The worms move along boundary of $D$ until they collide (Fig.~\ref{propagationatinterface}g)  and annihilate in the result of the collision (Fig.~\ref{propagationatinterface}h). The configuration of oscillating localizations then returns to its original state (Fig.~\ref{propagationatinterface}i). 

\clearpage

\section{Building conductive pathways}
\label{pathways}

\begin{finding}
By exploring collisions between excitation wave-fragments travelling in $\mathcal{A}_2$ one can built
information transmission pathways in an initially non-conductive medium. Routing primitives realised
include signal splitting, signal echoing and signal turning. 
\end{finding}

\begin{figure}[!htb]
\centering
\subfigure[$t=1$]{\includegraphics[width=0.32\textwidth]{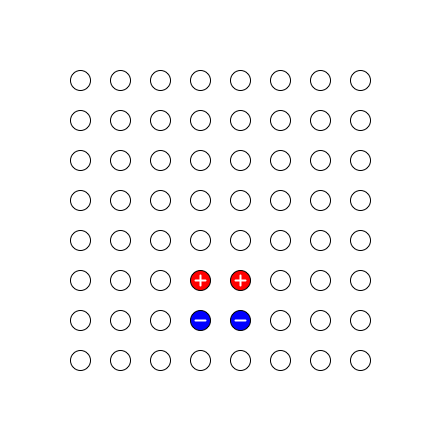}}
\subfigure[$t=1$]{\includegraphics[width=0.32\textwidth]{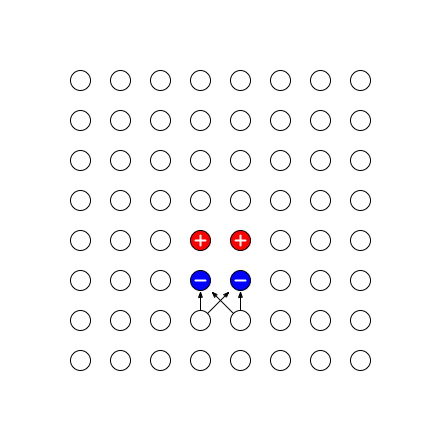}}
\subfigure[$t=1$]{\includegraphics[width=0.32\textwidth]{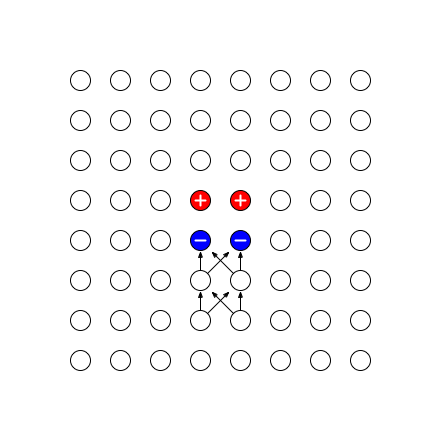}}
\subfigure[$t=1$]{\includegraphics[width=0.32\textwidth]{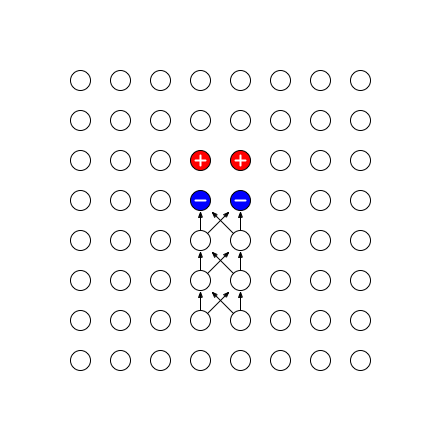}}
\subfigure[$t=1$]{\includegraphics[width=0.32\textwidth]{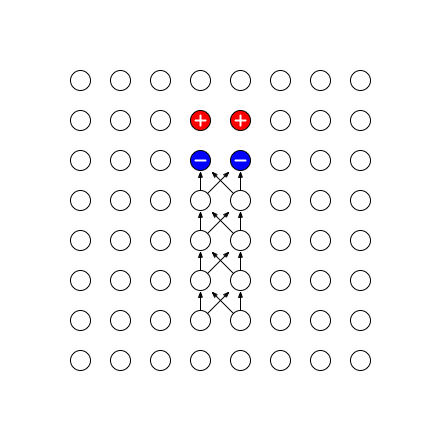}}
\subfigure[$t=1$]{\includegraphics[width=0.32\textwidth]{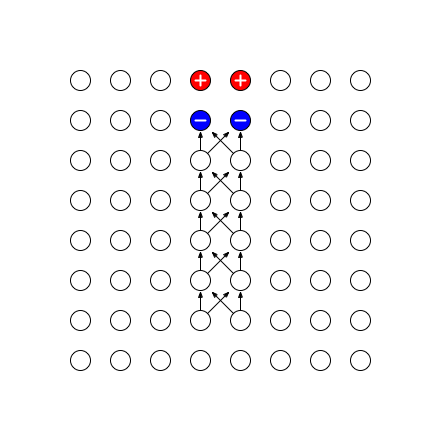}}
\caption{Snapshots of excitation and links dynamics in automaton  $\mathcal{A}_2^{10}$, 
$L_0$-start.  In initial configuration two cells are excited, two cells are in refractory state, 
others are resting.  Arrow from cell $x$ to cell $y$ means $l_{xy}=1$. }
\label{2PlusParticleToplogy}
\end{figure}

Localizations travelling in $\mathcal{A}_2$, $L_0$-condition, can form 
pathways conductive for low-strength excitations. For example, a localisation of two excited and two refractory states
propagates in $\mathcal{A}_2$ in the direction of its excited 'head'. The localisation forms a chain of conductive links 
oriented opposite to the localisation's velocity vector in case of $\mathcal{A}^{01}_2$, and in the direction of the localisation's propagation in case of $\mathcal{A}^{10}_2$ (Fig.~\ref{2PlusParticleToplogy}).

\begin{figure}[!htb]
\centering
\subfigure[$t=1$]{\includegraphics[width=0.32\textwidth]{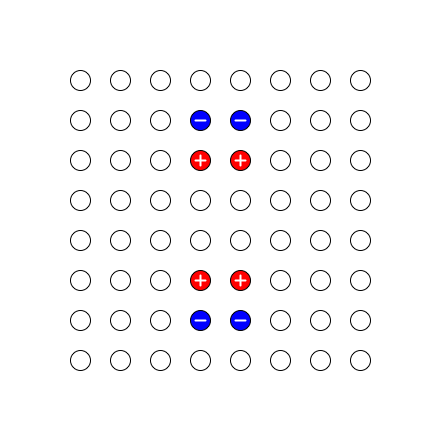}}
\subfigure[$t=1$]{\includegraphics[width=0.32\textwidth]{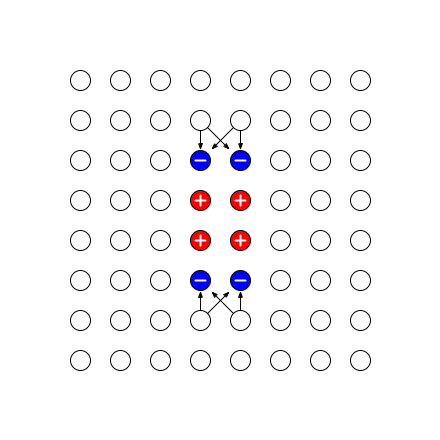}}
\subfigure[$t=1$]{\includegraphics[width=0.32\textwidth]{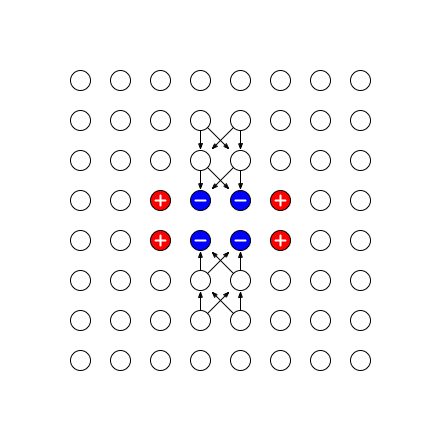}}
\subfigure[$t=1$]{\includegraphics[width=0.32\textwidth]{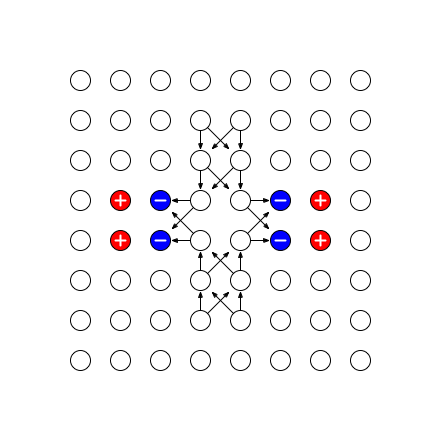}}
\subfigure[$t=1$]{\includegraphics[width=0.32\textwidth]{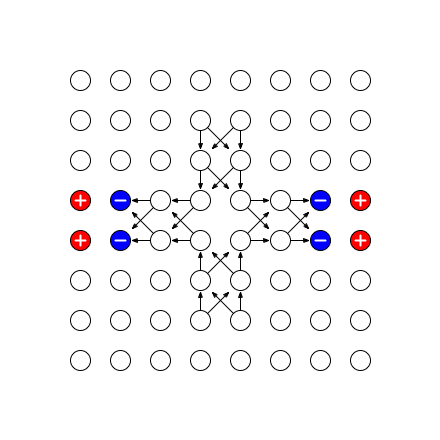}}\\
\subfigure[$t=1$]{\includegraphics[width=0.32\textwidth]{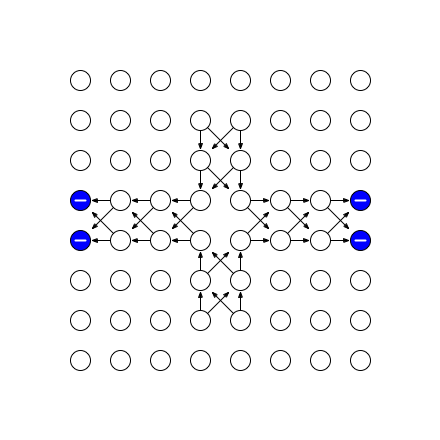}}
\subfigure[$t=1$]{\includegraphics[width=0.32\textwidth]{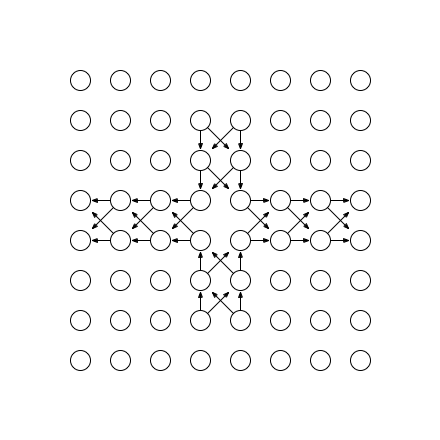}}
\caption{Snapshots of excitation and link dynamics in automaton  $\mathcal{A}_2^{10}$, 
$L_0$-start. In initial configuration two $2^+$-particles are introduced in the medium. 
Four cells are excited, four cells are in refractory state, others are resting. Arrow from cell $x$ to 
cell $y$ means $l_{xy}=1$. }
\label{2PlusParticleCollisionReflection}
\end{figure}

\begin{sidewaysfigure}[!htb]
\begin{tabular}{c|ccccc}
       &  0 & 1 & 2 & 3 & 4 \\ \hline
odd & \includegraphics[scale=0.3]{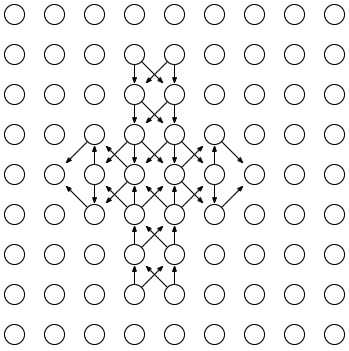} & \includegraphics[scale=0.3]{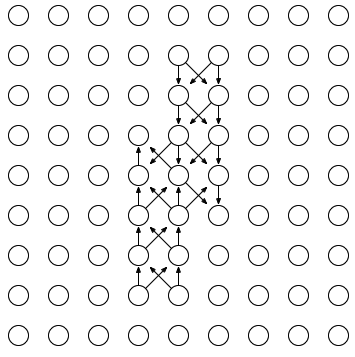}  &  \includegraphics[scale=0.3]{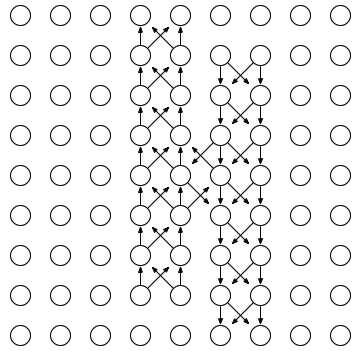}  &  \includegraphics[scale=0.3]{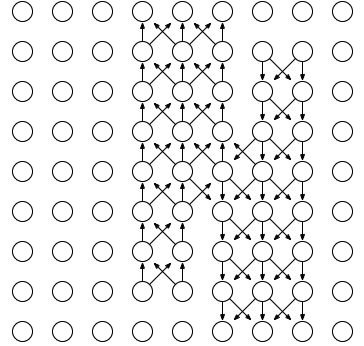} &  \includegraphics[scale=0.3]{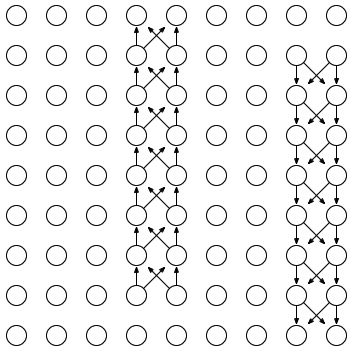} \\
even & \includegraphics[scale=0.3]{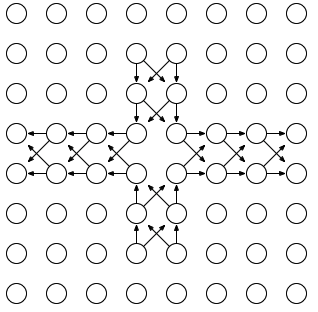} & \includegraphics[scale=0.3]{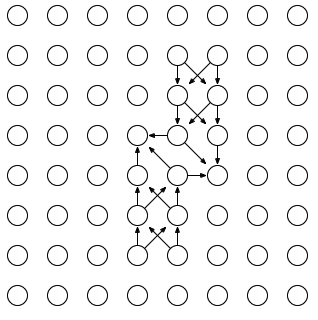} & \includegraphics[scale=0.3]{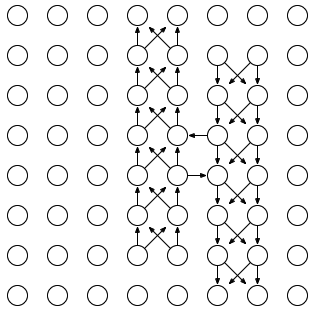} & \includegraphics[scale=0.3]{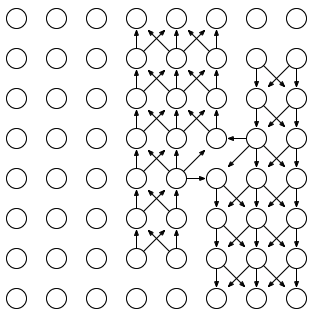} &  \includegraphics[scale=0.3]{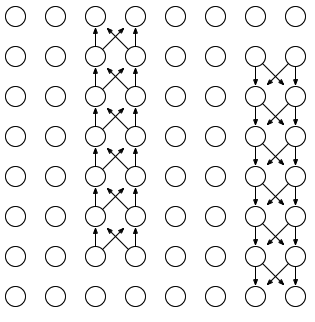}\\
\end{tabular}
\caption{Intersection of 'wires' build by propagating and colliding $2^+$-particles. 
Automaton $\mathcal{A}_2^{10}$.}
\label{2ParticlesInteractions}
\end{sidewaysfigure}

A layout of conductive pathways, or 'wires', is determined by outcomes of collisions between path-laying particles. A detailed example of such collision-determined pathway building is shown in Fig.~\ref{2PlusParticleCollisionReflection}. Two travelling localizations, $2^+$-particles\footnote{$2^+$-particles are localizations consisting of two excited and two refractory states, which move along rows or columns of an excitable cellular 
array~\cite{adamatzky:2001}},  are initiated in 
$\mathcal{A}_2^{10}$ facing each other with their excited heads (Fig.~\ref{2PlusParticleCollisionReflection}a). One particle propagates south another north  (Fig.~\ref{2PlusParticleCollisionReflection}a). When the particles collide they undergo elastic-like collision, in the result of which two $2^+$-particles are formed: one travels east another west 
(Fig.~\ref{2PlusParticleCollisionReflection}c--f).     When a weak excitation rule --- a cell is excited if at least one neighbour is excited and no links are updated --- is imposed on the automaton the pathways formed by these two colliding particles becomes selective. If an excitation is initiated in the southern or northern channel the excitation propagates till cross-junction and then branches into eastern and northern channels. 

Pathways built by two $2^+$-particles undergoing head-on collision being in different phases (odd and even 
distance between their start positions) and lateral offsets are shown in Fig.~\ref{2ParticlesInteractions}. Most
 T-bone collisions between $2^+$-particles lead to formation of omni-directionaly growing excitation patterns. Only in  situations when one particle hits a tail of another particle no uncontrollable growth occurs. The particle hitting a tail of another particle extinguishes and the other particle continues its journey undisturbed. 

\begin{figure}[!htb]
\centering
\subfigure[]{\includegraphics[scale=0.45]{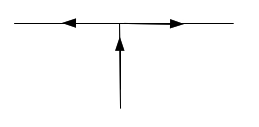}}
\subfigure[]{\includegraphics[scale=0.45]{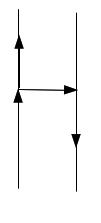}}
\subfigure[]{\includegraphics[scale=0.45]{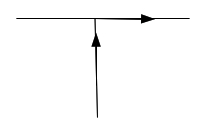}}
\caption{Types of information transfer pathways laid out by interacting travelling localisations:
(a)~T-branching, (b)~echo, (c)~turning.}
\label{routes}
\end{figure}

There are primitives of information routing implementable in collisions between $2^+$-particles 
(Fig.~\ref{routes}).  T-branching, or signal splitting,  (Fig.~\ref{routes}a)
 is built by particles colliding with zero lateral shift and even number of cells between their initial positions
 (e.g. Fig.~\ref{2ParticlesInteractions}, entry (even, 0)). When signal travelling north reaches cross-junctions it splits into two signals -- one travel west and another travels east; no signal continues straight propagation across junction. 
 
 Echo primitive (Fig.~\ref{routes}b) is constructed in a head-on collision between $2^+$-particles with lateral shift two or three cells  (see  (e.g. Fig.~\ref{2ParticlesInteractions}, entries (odd, 2), (odd, 3), (even, 2), (even, 3)).  The echo primitive consists  of two anti-aligned (e.g. one propagates information northward and another southward) parallel information pathways.  There is a bridge between the pathways. When signal propagating along one pathway reaches the bridge, it splits, daughter signal enters second pathway and propagates in the direction of mother signal's origination. In   (Fig.~\ref{routes}b) mother-signal propagates north and daughter-signal south. 
 
 Turn primitive  (Fig.~\ref{routes}c) is implemented in T-bone collision between two $2^+$-particles. A signal generated at either of the pathways propagates in the direction of $2^+$-particle which trajectory was undisturbed during collision.

\section{Discussion}
\label{discussion}

We  designed a minimalistic model of a two-dimensional discrete memristive medium. Every site of such medium takes triple states, and a binary conductivity of links is updated depending on states of sites the links connect. The model is a hybrid between classical excitable cellular automata~\cite{greenberg:1978} and classical structurally-dynamic cellular automata~\cite{ilachinsky:halpern:1987}.  A memristive automaton with binary cell-states would give us even more elegant model however by using binary cell-states we could not easily detect source and sink of simulated 'currents'.  Excitable cellular automata provide us with all necessary tools to imitate current polarity and to control local conductivity. From topology of excitation wave-fronts and wave-fragments we can even reconstruct relative location of a source of initiated current.

We defined two type of memristive cellular automata and characterised their space-time dynamics in response to point-wise and spatially extended perturbations. We classified several regimes of automata excitation activity, and provided detailed accounts of most common types of oscillating localizations.  We did not undertake any systematic search for minimal oscillators though but just exemplified two most commonly  found after random spatially-extended stimulation. Exhaustive search for all possible localised oscillations could be a topic of further studies.

\begin{figure}[!htb]
\centering
\subfigure[]{\includegraphics[width=0.59\textwidth]{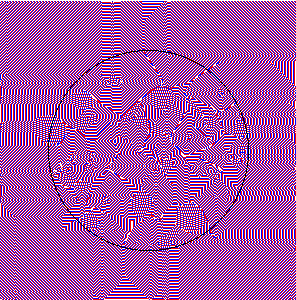}}
\subfigure[]{\includegraphics[width=0.59\textwidth]{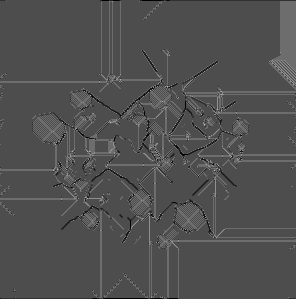}}
\caption{Configuration of target waves~(a) and configuration of cell in-degrees~(b) in $\mathcal{A}_2^{01}$,
$L_0$-start,  $D$-stimulation, $E_1$-excitation.  Boundary of $D$ is shown by circle in (a). }
\label{VDroutes}
\end{figure}

With regards to formation of conductive pathways just few possible versions amongst many implementable were discussed in the papers. Opportunities to grow 'wires' in memristive automata are virtually unlimited. For example,  
in $\mathcal{A}_2^{01}$, $L_0$-start, after first $D$-stimulation (Fig.~\ref{VDroutes}) generators of 
spiral and target waves are formed inside  $D$. Boundaries between the generators (they provide 
a partial approximation of a discrete Voronoi diagram over centres of the generators) are comprised of cells with 
high in-degrees. Such chains of high in-degree cells  can play a role of conductive pathways even if we increase excitation threshold of the medium.


\section*{Appendix}

\begin{figure}
\centering
\includegraphics[width=0.9\textwidth]{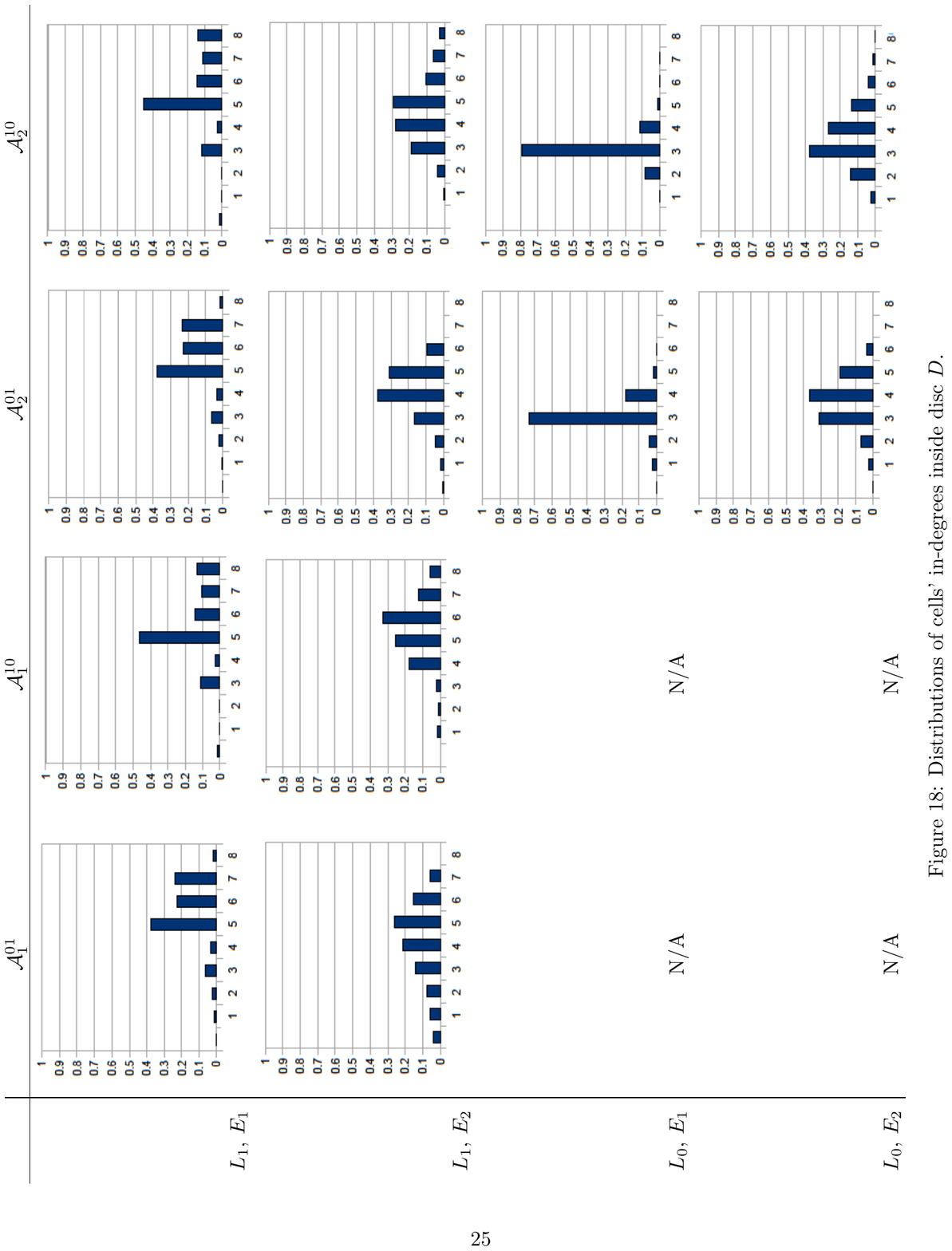}
\caption{Distributions of cells' in-degrees inside disc $D$. }
\label{insidediscdegrees}
\end{figure}

\begin{figure}
\centering
\includegraphics[width=0.9\textwidth]{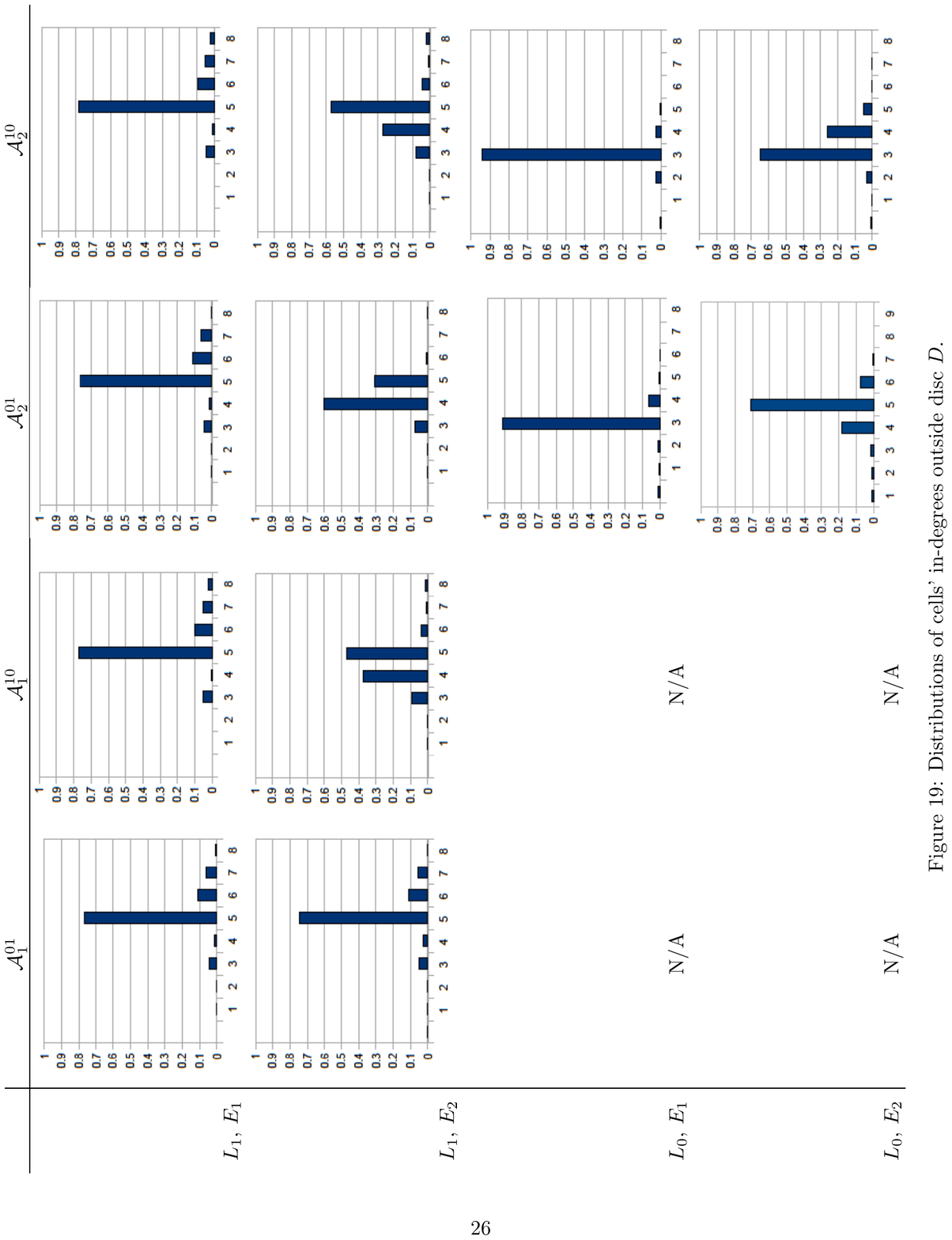}
\caption{Distributions of cells' in-degrees outside disc $D$.}
\label{outsidediscdegrees}
\end{figure}

Distributions of cell in-degrees for automata $\mathcal{A}_1$ and $\mathcal{A}_2$ after $E_1$- and $E_2$-excitations are shown in Figs.~\ref{insidediscdegrees} and ~\ref{outsidediscdegrees}. Calculations are done separately for cells lyings inside stimulation disc $D$ (Fig.~\ref{insidediscdegrees}) and outside the 
disc (Fig.~\ref{outsidediscdegrees}).  Each chart represents distribution of cell in-degrees, where horizontal axis is a number of incoming links and vertical axis is a ratio of cells with given in-degree to a total number of cells in the analysed domain. 

\clearpage

\clearpage

\begin{thebibliography}{99}


\bibitem
{adamatzky:1995}
Adamatzky~A. 
Controllable transmission of information in excitable medium: the $2^+$ medium.
Advanced Materials for Optics and Electronics 5 (1995) 145Ð-155.

\bibitem
{adamatzky:2001}
Adamatzky~A. 
Computing in Nonlinear Media and Automata Collectives (IoP, 2011).

\bibitem
{alonso-sanz:2006}
R. Alonso-Sanz
A structurally dynamic cellular automaton with memory,
Chaos, Solitons and Fractals 32 (2006) 1285Ð-1295.

\bibitem
{chua:1971}
 Chua~L.~O., Memristor --- the missing circuit element.
 IEEE Trans. Circuit Theory 18 (1971) 507--519.
 
 \bibitem
 {chua:1976}
 Chua~L.~O.  and Kang~S.~M., 
 Memristive devices and systems. 
 Proc. IEEE 64 (1976) 209--223.
 
 \bibitem
 {chua:1980} 
Chua~L.~O. 
Device modeling via non-linear circuit elements.
 IEEE Trans. Circuits Systems 27 (1980) 1014--1044.

\bibitem
{erokhin:2008}
 Erokhin V., Fontana M.T. Electrochemically controlled polymeric device: a memristors (and more) found two years ago. (2008) arXiv:0807.0333v1 [cond-mat.soft]



\bibitem
{halpern:1990}
Halpern~P. and G. Caltagirone.
Behavior of topological cellular automata.
Complex Systems 4 (1990) 623Ð651.



\bibitem
{hasslacher:1994}
Hasslacher~B. and Meyer~D.~A.
Modelling dynamical geometry with lattice gas automata.
Int J Modern Physics C 9 (1998) 1597--1605.


\bibitem
{hillman:1998}
Hillman~D.
Combinatorial spacetimes.
(Ph.D. dissertation) 234 pp.
\url{arXiv:hep-th/9805066v1}


\bibitem
{ilachinsky:halpern:1987}
Ilachinsky~A. and Halpern~P., Structurally dynamic cellular automata,
Complex Systems 1 (1987) 503--527.

\bibitem
{ilachinsky:2009}
Ilachinsky~A., 
Structurally Dynamic Cellular Automata
Encyclopedia of Complexity and Systems Science
2009, Part 19, 8815--8850.

\bibitem
{itoh:2009}
Itoh~M. and Chua~L. 
Memristor cellular automata and memristor discrete-time cellular neural networks.
Int. J. Bifurcation and Chaos 19 (2009) 3605--3656. 


\bibitem
{greenberg:1978}
Greenberg~J. M.  and Hastings~S. P.  
Spatial patterns for discrete models of diffusion in excitable media, 
SIAM J. Appl. Math. 34 (1978) 515Ð-523.

\bibitem
{requardt:2003}
Requardt~M.
A geometric renormalization group in discrete quantum spaceÐtime.
J. Math. Phys. 44 (2003) 5588.


\bibitem
{rose:1994}
Ros\'{e} H., Hempel~H., Schimansky-Geier~L. 
Stochastic dynamics of catalytic CO oxidation on Pt(100)
Physica A 206 (1994) 421--440. 


\bibitem
{strukov:2008}
Strukov, D.B., Snider, G. S., Stewart, D. R. and Williams, R. S., The missing memristor found. Nature 453 (2008) 80--83.


\bibitem
{tomita:2009}
Tomita~K., Kurokawa~H. and Murata~S.
Graph-rewriting automata as a natural extension of cellular automata. In: 
Understanding Complex Systems (Springer, 2009) 291--309.

\bibitem
{williams:2008}
Williams R.S. 
How we found the missing memristor. IEEE Spectrum 
2008-12-18.


\bibitem
{yang:2008}
Yang, J.J., Pickett, M. D., Li, X., Ohlberg, D. A. A., Stewart, D. R. and Williams, R.S.  Memristive switching mechanism for metal-oxide-metal nanodevices. Nature Nano, 2008 3(7). 

\end{thebibliography}
\end{document}